# MULTICHANNEL AUDIO SIGNAL SOURCE SEPARATION BASED ON AN INTERCHANNEL LOUDNESS VECTOR SUM

*Taejin Park and Taejin Lee*

Electronics and Telecommunications Research Institute, Republic of Korea

## ABSTRACT

In this paper, a Blind Source Separation (BSS) algorithm for multichannel audio contents is proposed. Unlike common BSS algorithms targeting stereo audio contents or microphone array signals, our technique is targeted at multichannel audio such as 5.1 and 7.1ch audio. Since most multichannel audio object sources are panned using the Inter-channel Loudness Difference (ILD), we employ the ILVS (Inter-channel Loudness Vector Sum) concept to cluster common signals (such as background music) from each channel. After separating the common signals from each channel, we employ an Expectation Maximization (EM) algorithm with a von-Mises distribution to successfully classify the clustering of sound source objects and separate the audio signals from the original mixture. Our proposed method can therefore separate common audio signals and object source signals from multiple channels with reasonable quality. Our multichannel audio content separation technique can be applied to an upmix system or a cinema audio system requiring multichannel audio source separation.

## 1. INTRODUCTION

In this paper, we tried to separate object and background signals from a multichannel audio source (such as 5.1ch, 7.1ch, or a higher channel format). The issue of audio Blind Source Separation (BSS) has been widely investigated using various approaches. The Degenerate Unmixing Estimation Technique (DUET) [1] is a source separation technique used to separate a source with an amplitude and phase difference. In addition, source separation for an underdetermined source using a Gaussian Mixture Model (GMM) [2] or Laplacian Mixture Model (LMM) was proposed [3]. Recently, BSS techniques for a multichannel microphone [4] and Multichannel audio content [5] have also been proposed.

However, unlike previous works related to BSS, which cluster signals using the ratio information from two channels [2,3,5], our proposed technique clusters an audio source from a multichannel signal at the same time as the vector concept. In addition, rather than taking inter-channel phase difference (IPD) into consideration such as in [1] and [4] for a multichannel microphone, we focus solely on ILD information since most multichannel movie audio tracks and multichannel music contents are mixed using the inter-channel loudness difference (ILD). We also employed the inter-channel loudness vector sum (ILVS) concept and set the channel loudness vector axis to express the panning information and uniqueness of the signal.

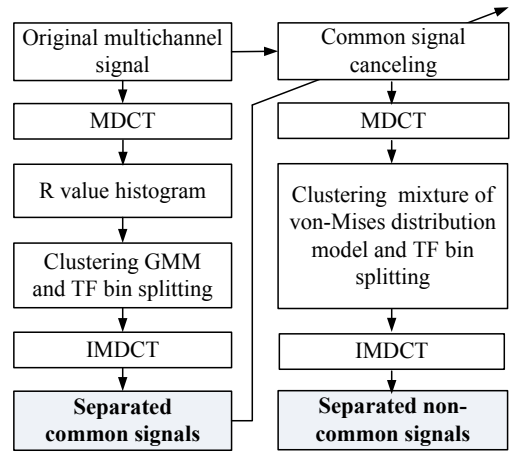

**Fig. 1.** Data flow of the proposed method.

Using the ILVS ratio, we are able to separate common signals such as background music from a multichannel audio signal. After separating common signals from an original multichannel source, we employ an EM algorithm with a von-Mises distribution, which is appropriate for a circular distribution, to cluster the sound object from the ILD. We therefore obtain separate sound object signals from multiple channels. Finally, we evaluated the performance of our algorithm using a well-known BSS evaluation technique [6].

## 2. VECTOR REPRESENTATION

As mentioned previously, we focused on the fact that most multichannel audio content expresses spatial sound information using the ILD. To analyze the panning information more properly through a vector representation, we set the axis for each channel with an equal interval angle, as shown in Fig. 2. The center channel is omitted because it typically contains only the speech signal. Like other previous works [1,2,3], the sparsity of the source is assumed, which means there is only one source signal TF (Time-Frequency) bin. Equation (1) describes how an ILVS for one TF bin is plotted on the vector axis. Equation (1) also describes how the ILVS and channel axis work. $|X_i(f,\tau)|$ is the absolute value of the Modified Discrete Cosine Transform (MDCT) of the i-th channel signal at the TF bin $(f,\tau)$, where $f$ indicates the frequency and $\tau$ is the frame index. These values are multiplied by the axis rotation matrix with channel axis angle $\theta_i = i \cdot \theta_v$. Thus, the ILVS value $\mathbf{V}(f,\tau)$ equals the sum of the angular transformed signal vectors, $\mathbf{s}_i(f,\tau)$, from each channel.

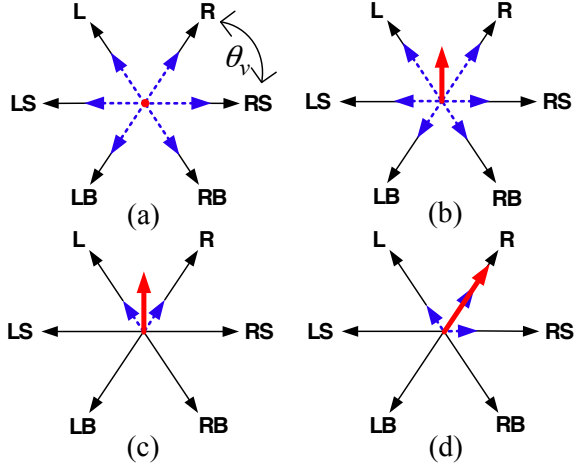

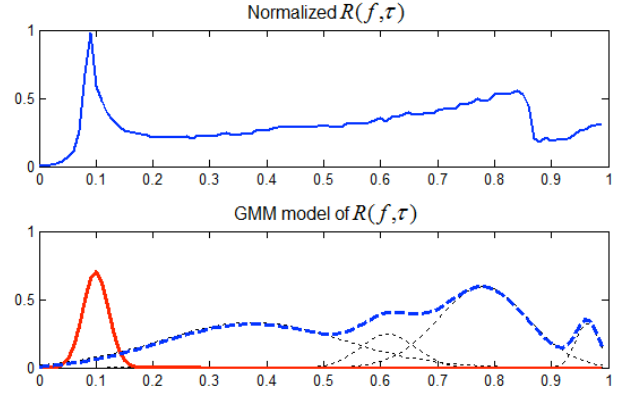

**Fig. 2.** Example vector representation and the ILVS. The loudness of each channel signal is described with a dashed arrow, and the ILVS is described with a solid arrow: (a) common signal ILVS with equal loudness on every channel, (b) common signal ILVS with biased loudness, (c) non-common ILVS of two-channel panning, and (d) non-common ILVS of three-channel panning.

$$\mathbf{s}_i(f,\tau) = \begin{bmatrix} |X_i(f,\tau)| \\ 0 \end{bmatrix}$$
$$\mathbf{T}_i(f,\tau) = \begin{bmatrix} \cos(\theta_i) & -\sin(\theta_i) \\ \sin(\theta_i) & \cos(\theta_i) \end{bmatrix} \quad (1)$$
$$\mathbf{V}(f,\tau) = \sum_{i=1}^{N} \mathbf{T}_i(f,\tau)\,\mathbf{s}_i(f,\tau) = \begin{bmatrix} X_i(f,\tau) \\ Y_i(f,\tau) \end{bmatrix}$$

We were therefore able to draw the ILVS from every TF bin and obtain clusters over a 360 degree polar plot, as shown in Fig. 4.

### 3. COMMON SIGNAL SEPARATION

Unlike stereo audio content, in multichannel audio, a common signal such as background music can exist on more than two channels. For a multichannel environment, conventional two-channel ratio-based clustering [2,3,5] is unable to distinguish a common signal from a center-panned signal. This problem is easily described in Figs. 2 (b) and 2 (c). If we only monitor two channels, a non-common signal that only exists on channels L and R, and a common signal that exists on every channel, can be clustered in the same position, as shown in Figs. 2 (b) and 2 (c). To solve this problem, we defined variable $R(f,\tau)$, which is the ratio of the sum of the individual loudness of the channel signal and ILVS. $R(f,\tau)$ can be described as follows.

**Fig. 3.** Plot of $R(f,\tau)$ value in equation (2), where the angular interval of the channel axis is $\theta_v = \pi/3$ in 7.1ch audio.

$$R(f,\tau) = \frac{|\mathbf{V}(f,\tau)|}{\sum_{i=1}^{N}|X_i(f,\tau)|} \quad (2)$$

$R(f,\tau)$ is determined based on the number of channels involved in the ILVS. If we set the channel axis interval as $\theta_v$ and assume an equal loudness for the channels involved, the ideal value of $R(f,\tau)$ can be determined as shown below.

$$\begin{array}{ll} \text{One channel :} & R(f,\tau) = 1 \\ \text{Two channel :} & R(f,\tau) = \cos(\theta_v/2) \\ \text{Three channel :} & R(f,\tau) = (1+\cos(\theta_v))/3 \\ \text{All channel :} & R(f,\tau) = 0 \end{array} \quad (3)$$

If we plot $R(f,\tau)$ from every TF bin in the original multichannel signal, the $R(f,\tau)$ value will be clustered as shown in the upper section of Fig. 3. In a real-life situation, the peaks are not exactly placed as predicted by equation (3) because the loudness of the involved channel signal is not equal. However, for a common signal, if one exists, the peak tends to be located near zero, which is far from the other peaks. We applied the EM algorithm with GMM to identify the leftmost peak location and width. The lower area in Fig. 4 shows the leftmost peak and the other part modeled using the GMM. After obtaining the $R(f,\tau)$ value of the leftmost peak, we are able to separate a common signal from a mixture of the source for each channel based on a sparsity assumption. Equation (4) describes the classification criterion of a common signal where

$$\begin{cases} X_{c,i}(f,\tau) = X_i(f,\tau) & \text{if } |r(f,\tau) - \mu_r| \leq d \\ X_{c,i}(f,\tau) = 0 & \text{if } |r(f,\tau) - \mu_r| > d \end{cases} \quad (4)$$

$\mu_r$ denotes the mean value of the leftmost Gaussian model, and $d$ is the decision threshold. $X_{c,i}(f,\tau)$ and $X_{n,i}(f,\tau)$ are common and uncommon signals of the i-th channel, respectively. After separating the common signals from each channel, we constructed the original common signal by taking the average of all common signals from all channels to minimize the interference from non-common signals, as described in (5). After obtaining the sum of the common signals, we subtracted the common signals from the original signal. During subtraction, we adjusted the gain using the root mean square (RMS) ratio of the averaged common signal and individual common signal from the i-th channel, as described in equation (6), where

$$X_{c,avg}(f,\tau) = \frac{1}{N} \sum_{i=1}^{N} X_{c,i}(f,\tau) \quad (5)$$

$$x_{n,i}(t) = x_{org,i}(t) - \left( \frac{RMS(x_{c,i})}{RMS(x_{c,avg})} \right) x_{c,avg}(t) \quad (6)$$

$x_{n,i}(t)$, $x_{org,i}(t)$, $x_{total,i}(t)$ are the desired non-common signal in the time domain, original i-th channel signal in the time domain, and time-domain signal of $X_{c,total}(f,\tau)$ in equation (5), respectively. Since a common signal is usually a wideband music signal in practice, common signals tend to overlap with non-common signals in the TF bin. Subtraction in the time domain can significantly reduce artifacts caused by the TF bin overlapping when compared to separation using a sparsity assumption.

## 4. NON-COMMON SIGNAL SEPARATION

Since we should express our signal in a polar coordinate, we had to consider the ILVS in the polar coordinate to have circularity. Therefore, if we apply a traditional EM method using a GMM [2] or Laplacian model [3] to a polar coordinate, it will suffer from an edge effect since Gaussian and Laplacian distributions have no circularity and are limited to both ends. Despite the existence of techniques to compensate the edge effect, the proposed compensation technique cannot solve the edge effect perfectly. To deal with this problem, we applied a von-Mises distribution [7] using the EM algorithm. Since a von-Mises distribution is based on a circular domain, this approach never suffers from an edge effect. Fig. 4 (b) shows the result of the clustering.

### 4.1. EM procedure with the von-Mises distribution.

If we express the polar coordinates as Cartesian coordinates, the 2-D variables $(|X_i(f,\tau)|, |Y_i(f,\tau)|)$ in equation (1) and the mean values of those variables $(\mu_x, \mu_y)$ in the Cartesian coordinates can be described as follows:

$$|X_i(f,\tau)| = r\cos(\theta), \quad |Y_i(f,\tau)| = r\sin(\theta) \quad (7)$$
$$\mu_x = r_0 \cos(\theta_0), \quad \mu_y = r_0 \cos(\theta_0)$$

where $r_0$ and $\theta_0$ are the mean radius and mean angle value, respectively. To project all values on the unit circle, we set $r = 1$. Therefore, after substituting (5) into the Gaussian distribution, the von-Mises distribution can be expressed as follows:

$$f_v(\theta | \theta_0, m) = \frac{1}{2\pi I_0(m)} \exp\{m\cos(\theta - \theta_0)\} \quad (8)$$

where $I_0$ is the Bessel function of the first kind, and $m = r_0/\sigma^2$, where $\sigma$ is the variance of the original Gaussian distribution. With this distribution, we employ the EM algorithm to every TF bin data and express the total distribution with a mixture of von-Mises distributions. The E and M steps of the applied EM algorithm can be described as follows.

1) **E-step**: Evaluate the responsibility $\gamma_{nk}$ with weight $\pi_k$ and von-Mises distribution function $f_v(\theta_n | \theta_{0,k}, m_k)$, where $k$ denotes the index of the von-Mises distribution model, and $n$ denotes the sample index.

$$\gamma_{nk} = \frac{\pi_k f_v(\theta_n | \theta_{0,k}, m_k)}{\sum_{j=1}^{M} \pi_j f_v(\theta_n | \theta_{0,k}, m_k)} \quad (9)$$

2) **M-step**: Estimate the weight values $\theta_{0,k}$ and $m_k$ using a maximum likelihood estimator. Evaluate the weight value from responsibility $\gamma_{nk}$

$$\pi_k = \frac{1}{N} \sum_{n=1}^{N} \gamma_{nk}$$

$$\theta_{0,k} = \tan^{-1} \left( \frac{\sum_{n=1}^{N} \gamma_{nk} \sin(\theta_n)}{\sum_{n=1}^{N} \gamma_{nk} \cos(\theta_n)} \right) \quad (10)$$

$$A(m_k) = \frac{I_1(m_k)}{I_0(m_k)} = \frac{\sum_{n=1}^{N} \gamma_{nk} \cos(\theta_n - \theta_{0,k})}{\sum_{n=1}^{N} \gamma_{nk}}$$

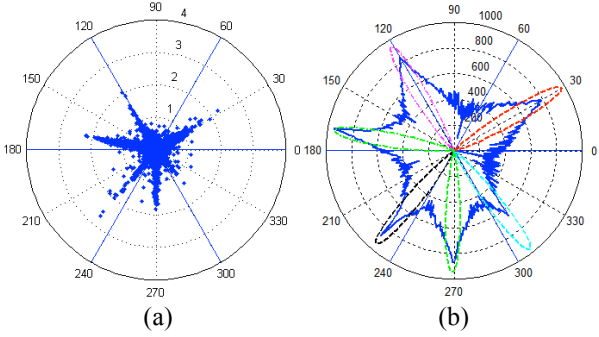

**Fig. 4.** (a) Plot of $\mathbf{V}(f,\tau)$ values in polar coordinates. (b) Plot of log scale histogram of (a) and result of clustering with mixture of von-Mises distribution

---

**Algorithm. 1.** Channel signal summation decision rule

**Inputs :** $\mathbf{V}(f,\tau)$: ILVS     $R(f,\tau)$ : R value
$X_i(f,\tau) = X_{n,i}(f,\tau)$: i-th channel signal (non-common)
**Outputs :** $S_k(f,\tau)$ : k-th separation output
1: **if** $\mathbf{V}(f,\tau)$ is on i-th channel axis
2:     **if** $R(f,\tau) \approx 1$ **do** $S_k(f,\tau) = X_i(f,\tau)$
3:     **else do** $S_k(f,\tau) = X_{i-1}(f,\tau) + X_i(f,\tau) + X_{i+1}(f,\tau)$
4: **if** $\mathbf{V}(f,\tau)$ is between i-th channel and (i-1)-th channel axis **do** $S_k(f,\tau) = X_i(f,\tau) + X_{i-1}(f,\tau)$
5: **if** $\mathbf{V}(f,\tau)$ is between i-th channel and (i+1)-th channel axis **do** $S_k(f,\tau) = X_i(f,\tau) + X_{i+1}(f,\tau)$
7: **return** $S_k(f,\tau)$

---

### 4.2. Separation of an object signal from a channel signal

A histogram of the ILVS data can be clustered as shown in Fig. 5 using the EM algorithm described in 5.1. If the data are clustered properly, the sound source can be separated based on the panning location. The decision rule (Algorithm 1) shows how we extract a signal from the original channel signal. If the ILVS is on the i-th axis, it may be the result of multiple (for most cases, three) channels or a single channel. These cases can be distinguished by the $R(f,\tau)$ value in equation (2). If the ILVS is not on the channel axis, it must be the result of two or more channels. The nearest two channels would be the most probable case. If there are more than three channels, the other one or two channel signals will be very weak, and even if they are loud, will be considered as common signals.

### 5. EXPERIMENTAL RESULTS

To test the performance of our algorithm, we made our test set using background music as a common signal, and sound effects from movies as a non-common signal. The overall performance was evaluated using the algorithm proposed in [6]. The experimental results of our proposed technique are described in the table 1 and table 2. We tested our proposed algorithm with different panning combination. "n-ch panning source" in the Table 1 and Table 2 means that n-channel is involved to pan one sound source. Four sources were used for test 5.1ch system and six sources were used for test 7.1ch system. The performances of non-common signal sources were averaged. Since proposed common signal separation technique uses common signal which exists in every channel and take average of extracted common signals, common signal showed better source separation quality than non-common signals.

|  |  | mixA | mixB | mixC | mixD |
|---|---|---|---|---|---|
| Number of 1ch panning source | | 2 | 0 | 0 | 0 |
| Number of 2ch panning source | | 2 | 4 | 3 | 2 |
| Number of 3ch panning source | | 0 | 0 | 1 | 2 |
| Number of common signal source | | 1 | 1 | 1 | 1 |
| Non-common signal (dB, Averaged) | SAR | 9.4 | 10.0 | 8.7 | 6.1 |
| | SDR | 9.1 | 9.7 | 8.4 | 4.9 |
| | SIR | 24.1 | 22.4 | 21.2 | 15.8 |
| Common signal (dB) | SAR | 11.9 | 11.6 | 12.2 | 11.9 |
| | SDR | 11.7 | 11.4 | 12.0 | 9.6 |
| | SIR | 26.4 | 25.6 | 25.7 | 13.6 |

**Table. 1.** Performance measurement of 5.1ch multichannel format.

|  |  | mixA | mixB | mixC | mixD |
|---|---|---|---|---|---|
| number of 1ch panning source | | 3 | 0 | 0 | 0 |
| number of 2ch panning source | | 3 | 6 | 5 | 4 |
| number of 3ch panning source | | 0 | 0 | 1 | 2 |
| Number of common signal source | | 1 | 1 | 1 | 1 |
| Non-common signal (dB, Averaged) | SAR | 3.0 | 3.5 | 3.5 | 2.8 |
| | SDR | 2.5 | 2.8 | 2.9 | 1.8 |
| | SIR | 14.6 | 16.0 | 14.8 | 12.9 |
| Common signal (dB) | SAR | 10 | 10.5 | 10.2 | 9.7 |
| | SDR | 9.8 | 10.3 | 9.9 | 9.5 |
| | SIR | 23.2 | 23.7 | 22.7 | 23.0 |

**Table. 2.** Performance measurement of 7.1ch multichannel format.

### 6. CONCLUSION

We proposed a technique to separate sound sources from multichannel audio panned using the ILD. By employing a vector representation, we successfully separated a common signal and non-common signal with reasonable quality. This result can be applied to an up-mix algorithm that needs to analyze spatial information and extract an object audio signal from multichannel audio content, such as that proposed by [5]. In addition, using our proposed technique, it is possible to control the volume of the background music and other sound effects separately while watching a movie or listening to multichannel audio content.